# INTERFEROMETRY AND HOLOGRAPHY WITH DIODE LASER LIGHT

José Joaquín Lunazzi, Wendel Lopes Moreira, Campinas State University-UNICAMP-IFGW
C.P.6165   13084-100Campinas - SP - Brazil    lunazzi@esperanto.nu

**Abstract**
   We made an interferometric Michelson type setup and a simple holographic setup to demonstrate the feasibility of interferometric and holographic techniques by means of a diode laser. The laser was made by using a common diode available as a penlight element  (less than R$ 15,00 value) and a simple stabilized 110 VCA- 3 VCC power supply. Interference fringes and holograms of small objects where obtained very similar to those  of a helium-neon laser based setup.

**Resumo** (esperanto)
   Ni faris Michelson interferometron kaj facila holografia sistemo por pruvi la eblecon fari interferometriajn kaj holografiajn teknikojn per lasera diodo. La lasera diodo ni faris per ordinara diodo achetebla kiel lanterno (malpli ol R$ 15,00 kosta) kaj ordinara stabiligita 110 VAK- 3 VKK elektrika  fonto. Interferencajn franghojn kaj hologramojn de malgrandaj objektoj ni obtenis tre similaj al la faritaj per heliumo-neona lasera sistemo.

**Introduction**
    The appearance on the market of very cheap, small and portable penlight laser diodes is a marvelous surprise for researches who deals with laser light since  more than 30 years, always waiting for the laser costing less than u$ 100,00 to appear. The change was sudden, and of one order of magnitude in the price, weigt and size, thanks to the massive production and distribution made by oriental countries, namely China. Recognising the beam so similar to our familiar red beam of helium-neon lasers our first intention was to profite it for educational purposes.  As we did not knew the coherence length of this laser junction and were affraid to be reduced due to a large bandwidth, our first approach was to employ it for viewing and projecting holograms. We employed the same holograms we already made for selling to teachers and schools **(1)** , and projected the real image on a screen.  The result was entirely equivalent to that of the projection with helium-neon laser red light, we then made the virtual image by adding a diverging lens to the system, and were satisfied with the possibility of divulgating holography with a setup where the less expensive component now is the penlight laser.
We knew that single mode emission is necessary for a long coherence length, and that it is obtained in infrared diode lasers by means of filtering at the cavity, adding a Fabry-Perot cavity or a Distributed Feedback cavity (DFB). So that we were skeptic about the possibility of making holograms with those lasers, and no news indicated that possibility.  A previus attempt reported to us **(2)** consisted of trying a Denisyuk simple single-beam hologram and was almost unsuccesful and one could only see a very small part of the objectr (a coin). Latter in december 1998 we could see a demonstration by means of a home-made Michelson interferometer **(3)** consisting of ordinary glass second surface mirrors and wood supports, and the reported easy of alignment made us think that coherence length should be larger than many microns, about one milimeter or more, to allow such experience. So that, after succeeding in making a simple regulated power supply where we mounted the head of a penlight laser, measuring a power of more than 2 mW for its emission, and obtaining good Lippmann interference on holographic film we decided to try a Michelson interferometer and a holographic setup. At the same time, we knew from informal reports that holography using similar lasers was obtained by holographers of USA, who reported coherence lengths of some meters.  It is interesting to know that such important developments are not widely known on the academic community and that we could not find publications on the subject, maybe due to the ordinary delay publications have.

**Previous experiences**
Lippmann interference was obtained by impinging the laser beam on a 2 mm diameter surface of AGFA 8E 75 holographic film, having the emulsion side opposite to the laser and faced to the aluminized reflective surface of a plastic sheet. After developing and bleaching like in an ordinary Denisyuk holographic procedure we inmpinged white light at the surface, observing green reflected light. The green color was explained considering the shrinking of the holographic emulsion, usual in Denisyuk holography.

**Interferometry**
An interferometer was made by using two first surface aluminized mirrors mounted on angularly controled mounts and a not aluminized beam splitter mounted on a fixed mount. Ordinary  3 cm x 4 cm floating glass was the substrate for the mirrors and the material for the beam splitter.  The schematic setup is shown in Figure 1.

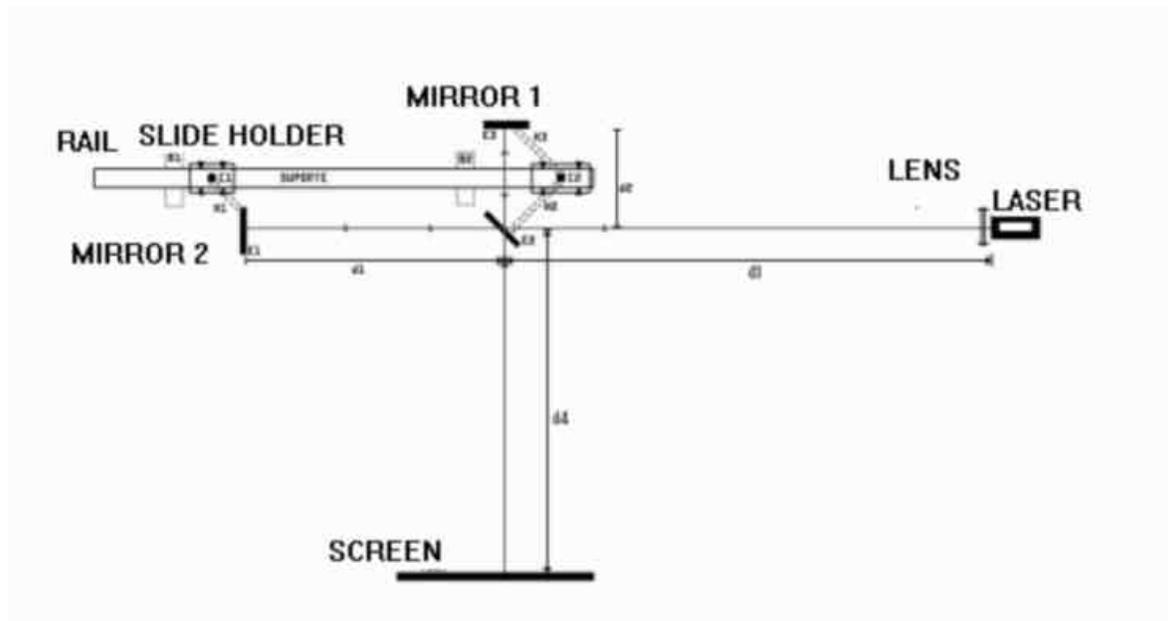

**Figure 1: Michelson interferometer setup**
The laser was at 80 cm from the beam splitter, the screen at 20 cm from it, and the mirrors could be at the equal path situation (10±0,2 cm for each arm) or one sliding mirror be displaced to give a path difference of 50±0,2 cm. The beam was seen to be uniform in intensity and rectangular in shape in the ratio 1 to 4. Different width of fringes were generated and the constrast was visually good, with no appreciable change when sliding the mirror, indicating a coherence lenght greater than a meter or so.

We see in Fig. 2 an example of the fringes.

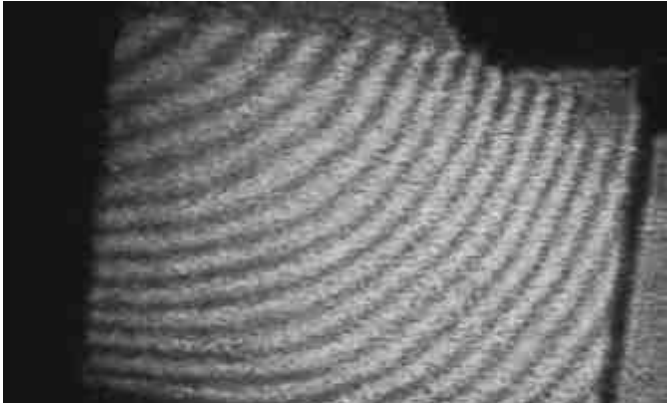

**Figure 2: Fringes obtained.**

Our system was not entirely stable until we inflated the pneumatic damping legs our granite optical table has.

**Holography**

Second, we made holograms by using a simple single-beam setup, as shown in Figure 3.

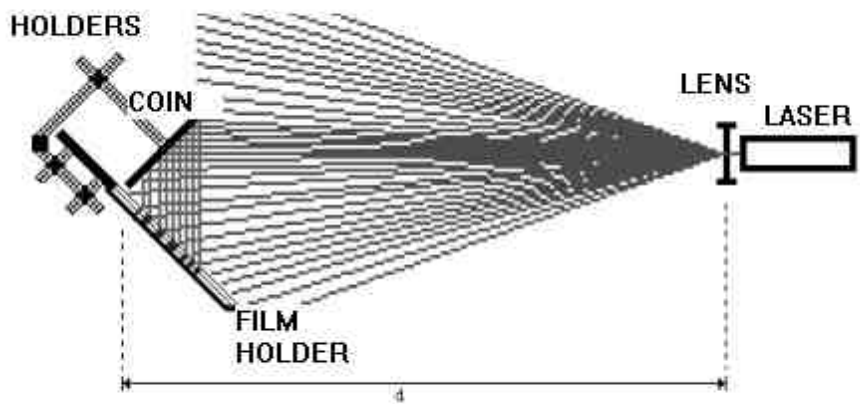

**Figure 3: Holographic setup**

This simple setup is very stable and we obtained 5 cm x 5 cm holograms of a simple object (a coin) using AGFA 8E75 holographic film holded between two glass plates.

A 5 min. waiting time was necessary for mechanical stabilisation of the film within the holder. As a result, we observed the holograms to appreciate that its visual quality is comparable to the holograms we are accostumed to make using helium-neon laser light.

This important result estimulated us to mount a 30-50 mW laser diode, which we are experiencing now.

**Conclusions**

It is demonstrated that interferometry and holography can be made with a simple diode laser. Although the technique was not evaluated with instruments, this conclusion is evident from our experience. It is to be noticed that we cannot assure that any ordinary laser diode will have the same performance, and that we had no information on the diode characteristics.

1) 10 cm x 12 cm laser hologram included on the demonstration setup employed since 1997 by Prof. Eduardo Cordeiro and co-workers, Physics Institute of Universidade Federal Fluminense, in lectures being diseminated through schools of the Rio de Janeiro State.
2) personal report and demonstration by Marcio M. Ueno, see also its Ph.D. thesis, São Paulo State University, Faculty of Education, 1997.
3) "Interferômetro de Michelson com um laser de caneta", work of a group of secondary school students at a science fair, oriented by graduate physics student J. Teles, exhibited at Institute of Mathematics, Campinas State University, Campinas-SP, Brazil, december 1998.